\documentclass[pra,a4paper,aps,twocolumn,showpacs,superscriptaddress,groupedaddress]{revtex4}
\usepackage[dvips]{graphicx}
\usepackage{ae}
\usepackage[T1]{fontenc}
\usepackage[ansinew]{inputenc}
\usepackage{amsmath}
\usepackage{amsthm}
\usepackage{amssymb}
\usepackage{graphicx}
\usepackage{caption}
\usepackage{color}
\usepackage[colorlinks]{hyperref}
\usepackage{lscape}

\begin{document}
\title{ Tight upper bound of genuine four party Svetlichny type nonlocality with and without local filtering  }
\author{Sk Sahadat Hossain}
\email{sk.shappm2009@gmail.com}
\affiliation{Department of  Mathematics, Nabagram Hiralal Paul College, Hooghly-721246, West Bengal, India}
\author{Biswajit Paul}
\email{biswajitpaul4@gmail.com}
\affiliation{Dept.Of Mathematics, Balagarh Bijoy Krishna Mahavidyalaya, Balagarh, Hooghly-712501, West Bengal, India}
\author{Indrani Chattopadhyay}
\email{icappmath@caluniv.ac.in}
\affiliation{Department of Applied Mathematics, University of Calcutta, 92, A.P.C. Road, Kolkata-700009, India }
\author{Debasis Sarkar}
\email{dsarkar1x@gmail.com, dsappmath@caluniv.ac.in}
\affiliation{Department of Applied Mathematics, University of Calcutta, 92, A.P.C. Road, Kolkata-700009, India }
\begin{abstract}
Identifying the nonlocality of a multiparty quantum state is an important task in quantum mechanics. Seevinck and Svetlichny [Phys. Rev. Lett. 89, 060401 (2002)], and independently, Collins and co-workers [Phys. Rev. Lett. 88, 170405 (2002)] have generalized the tripartite notion of Svetlichny nonlocality to n-parties. Here we have developed a tight upper bound for genuine four party Svetlichny type nonlocality. The constraints on the quantum states for the tightness of the bound are also presented. The method enables us to provide necessary and sufficient conditions for violating the four qubit Svetlichny type inequality for several quantum states. The relations between the genuine multipartite entanglement and the maximal quantum value of the Seevinck and Svetlichny operators for pure four qubit  states are also discussed. Consequently, we have exhibited genuine four qubit hidden nonlocality under local filtering. Our result provides an effective and operational method for further study of multipartite quantum nonlocality.
\end{abstract}
\date{\today}
\maketitle
\section{Introduction}
The nonlocal nature of quantum correlations, incompatible with local hidden variable theory (LHVT) are displayed through violations of various Bell inequalities \cite{1,2,3}. In a hierarchy of possible manifestations of the quantum correlation of the world, nonlocality is perhaps the strongest
one \cite{3,4,5}. It is significantly different from the classical description of physical phenomena \cite{6}. Nonlocality has important foundational implications as well as it is a useful operational resource \cite{7}, which plays an essential role, e.g., in the implementation of secure quantum
key distribution \cite{8,9,10}, building quantum protocols to decrease communication complexity \cite{11,12}, etc.\\
In the multipartite scenario, the rich and complex nature of quantum nonlocality is less explored than its bipartite counterpart \cite{3,13,14,15,16,17,18,19}. In the tripartite scenario, Svetlichny inequality (SI) \cite{13} has provided sufficient criteria to reveal genuine three-way nonlocality, while the standard form of tripartite nonlocality ( nonlocal correlation present among any two parties, locally correlated with the rest) is displayed through the violation of Mermin inequality (MI) \cite{14}. Three qubit  Greenberger-Horne-Zeilinger (GHZ) and W class states violate such inequalities \cite{16,17}. These are essentially the unique witness of nonlocality when all the three parties perform dichotomic measurements on their respective subsystems \cite{18}. Meanwhile, a reassessment of tripartite nonlocality has produced a series of weaker inequalities, whose violation can reveal tripartite nonlocality even when the SI is not violated \cite{7,18}.\\
Now, Seevinck and Svetlichny, as well as Collins et al. \cite{19}, have developed a sufficient condition for genuine $n$-particle nonlocality, i.e., for $n$-particle entanglement that cannot be reduced to mixtures of states in which a smaller number of particles are entangled. Seevinck and Svetlichny derived $n$-particle Bell-type inequalities under the assumption of partial separability. States are called partially separable if the n-particle system is composed of subsystems that may be correlated (e.g., entangled) in some way but are uncorrelated w.r.t. each other. States that violate inequalities in \cite{19} are known as genuine $n$-party nonlocal. These inequalities are maximally violated by the $n$-particle GHZ states \cite{19}. It is worth mentioning that the standard form of four qubit nonlocality with two local settings for each party is revealed through the violation of Mermin-Ardehali-Belinskii-Klyshko (MABK) inequality \cite{14,20,21}, which is maximally violated by the four qubit GHZ state \cite{22}. On the contrary, four-particle nonlocality with restricted measurement settings for one or more parties has been introduced in \cite{23,24}. These inequalities are maximally violated by four qubit cluster state \cite{25} and $ \vert \chi \rangle $ state \cite{26} respectively. In this regard, various attempts to classify four qubit entanglement are presented in \cite{27,28,29}. Multipartite quantum correlations are useful resources for computation \cite{30}, simulation \cite{31}, and metrology \cite{32}, hence the study of genuine multipartite correlation is a field of recent attraction.\\
The maximum value of bipartite Bell-CHSH operator \cite{2} for any two qubit quantum state is given by Horodecki's criteria \cite{33}. Recently, in a tripartite scenario such as the maximum bound of SI for pure GHZ and W state was studied in \cite{17}, while in Ref. \cite{34} the authors have pointed out analytical and numerical prescriptions for detecting the maximum quantum bound of the SI for Gaussian states. The tight upper bound of an arbitrary three qubit entangled state for SI is presented in \cite{35}, whereas a tight upper bound for MI has been established in \cite{36}.\\
In parallel to that, there exist entangled local states \cite{37} whose nonlocality can be revealed by local filtering operations \cite{38}, such traits are termed as "hidden nonlocality" \cite{39}. In \cite{40,41}, the authors consider two-qubit states that do not violate CHSH inequality initially, but do violate after performing local filtering operations. The maximal violation of the CHSH inequality and the lower bound of the maximal violation of the Vertesi inequality under local filtering operations are computed analytically in \cite{42}. A demonstration of hidden steerability under local filtering has been introduced in \cite{43}. In a tripartite system, genuine and standard hidden nonlocality under local filtering is exhibited in Ref. \cite{44,45} respectively.\\
In this present work, we have responded to the question: given a four qubit state, how to check whether it demonstrates genuine four-particle nonlocality ( non partial separability ) or not? Meanwhile, the genuine four party nonlocality is exhibited through the violation of the four party Seevinck and Svetlichny inequality (SSI) \cite{19}. We have formulated a tight upper bound for the maximum quantum value of four party SSI, where the maximum is attained by the four party GHZ state. Moreover, we have provided the constraints on the quantum state for the tightness of the bound. Consequently, the sufficient and necessary condition of violating four party SSI for several quantum states is given, including the white and color noised GHZ states. Moreover, we have found the relationship between genuine entanglement and nonlocality of several four qubit pure states. Further, we have investigated the maximum quantum value of the Seevinck and Svetlichny (SS) operators \cite{19} under local filtering procedures. A tight upper bound for the maximal value of the SS operators after local filtering is obtained. We have presented relevant examples to illustrate the importance of local filtering operations in producing hidden nonlocality for four qubit systems.\\ 
 The paper is organized as follows: In Sec. II, we have discussed the four party SSI in brief and have established a tight upper bound for Seevinck and Svetlichny (SS) operators. In Sec. III we have developed nonlocality criteria of some pure four qubit states based on entanglement measure. In Sec. IV we have established tight upper bound under local filtering and revealed hidden nonlocality with suitable example, finally we sum up with a conclusion in Sec. V.  

\section{The Seevinck and Svetlichny operator and its tight bound} 
We start with a brief review of the multipartite SSI \citep{19}. Consider a four party quantum system, namely Alice (A), Bob (B), Charlie (C) and Dick (D) each perform dichotomic measurements on their respective subsystems, with possible outcomes $ \pm1 $. Let the observables  $ X  $ ( $ X = A,\,A';\,B,\,B';\,C,\,C';\, D,\,D' ) $ are of the form $ X = \overrightarrow{x}.\overrightarrow{\sigma} $ = $ \Sigma_{k} x_{k}\sigma_{k} $, where $ \overrightarrow{x} \in \lbrace \overrightarrow{a},\,\overrightarrow{a'},\overrightarrow{b},\overrightarrow{b'},\overrightarrow{c},\overrightarrow{c'},\overrightarrow{d},\overrightarrow{d'}\rbrace $ correspondingly, $ \sigma_{k} $ ( $ k = 1,\,2,\,3) $ are the Pauli matrices with $ \overrightarrow{\sigma} = (\sigma_{1},\,\sigma_{2},\,\sigma_{3}), $ and $ \overrightarrow{x} = (x_{1},\,x_{2},\,x_{3}) $ is a three-dimensional real unit vector.\\
The four particle Seevinck and Svetlichny (SS) operators \cite{19} are defined as follow:
\begin{equation}\label{b1}
\begin{split}
 S_{4} = [A \otimes B-A' \otimes B'] \otimes [(C-C') \otimes D- (C+C') \otimes D']\\
 -[A' \otimes B + A \otimes B'] \otimes [(C+C') \otimes D+(C-C') \otimes D'].
\end{split}
 \end{equation} 
For any four particle partially separable states $ \vert \Phi \rangle $ (i.e., the state composed of subsystems that may be correlated (entangled) among itself but the subsystems are uncorrelated w.r.t. each other), the SS operator $ S_{4} $ is bounded by \cite{19},
\begin{equation}\label{b2}
 \vert\langle \Phi\vert S_{4}\vert \Phi\rangle\vert \leq 8.
 \end{equation}
  
\textbf{Theorem 1.} For any four qubit quantum state $ \rho $, the maximum quantum value $ V(S_{4})$ of the SS operator $ S_{4}$ defined in \eqref{b1} is bounded by\\
\begin{equation}\label{v1}
 V(S_{4})= \max \vert \langle S_{4}\rangle_{\rho}\vert \leq 4\sqrt{2} \lambda_{\max}.
\end{equation}
Where $ \langle S_{4}\rangle_{\rho} = $ Tr$(S_{4}\rho)  $, $ \lambda_{\max} $ is the largest singular value of the correlation matrix $ M = [t_{kl,ij}] $, where the elements are given by, $ t_{ijkl} =  $ Tr[$ \rho (\sigma_{i} \otimes \sigma_{j} \otimes \sigma_{k} \otimes \sigma_{l}) $],  $ i,j,k,l =1,2,3 $.\\
In order to set up the argument, we first present the following result from \cite{34}.\\
\textbf{Lemma.} Given a rectangular matrix A of size $ m \times n $, and any vector $ \overrightarrow{x} \in R^{m} $ and $ \overrightarrow{y} \in R^{n} $, we have\\
\begin{equation}\label{b3}
  \vert \overrightarrow{x}^{T} A \overrightarrow{y} \vert \leq \lambda_{\max}\vert\overrightarrow{x}\vert \vert\overrightarrow{y}\vert,
    \end{equation}  
    where $ \lambda_{\max} $ is the largest singular value of matrix $A$. The bound is tight when $ \overrightarrow{x} $ and $ \overrightarrow{y} $ are the corresponding singular vectors of $ A $ with respect to $ \lambda_{\max} $ (see Appendix \eqref{a1} for proof).\\
For the proof of Theorem 1, see Appendix \eqref{a2}. \\    
 From Theorem 1, it is clear that to saturate the upper bound one can have $ \theta_{ab} = \pm \frac{\pi}{2} $, hence, by proper choice of measurement settings for Alice and Bob the upper bound can be achieved. Again from the Lemma, we have the resulting inequality saturates if the degeneracy of $ \lambda_{\max} $ is more than 1, and corresponding to such $ \lambda_{\max} $ there are two nine-dimensional singular vectors of the form  $ (\overrightarrow{a} \otimes\overrightarrow{b}-\overrightarrow{a'} \otimes\overrightarrow{b'}) $ and $ (\overrightarrow{a'} \otimes\overrightarrow{b}+\overrightarrow{a} \otimes\overrightarrow{b'}) $, respectively.\\ 
\textbf{Example 1.} We have considered the mixture of the white noise and the four qubit GHZ-class states, which is given by,\\
\begin{equation}\label{b5}
\rho = p \vert \psi_{\theta} \rangle \langle \psi_{\theta} \vert +\frac{1-p}{4} \vert 00\rangle \langle 00\vert \otimes I_{2}^{\otimes 2},
\end{equation}
where $  \vert \psi_{\theta} \rangle = \cos \theta \vert 0000\rangle + \sin \theta \vert 1111\rangle $,  $ 0\leq p\leq 1 $ and $ I_{2} $ is the identity matrix. For $ \theta = \frac{\pi}{4} $, the matrix M is of the following form,\\
\begin{equation}
M = 
\begin{pmatrix}

  p & 0 & 0 & 0 & -p & 0 & 0 & 0 & 0\\
 0 & -p & 0 & -p & 0 & 0 & 0 & 0 & 0\\
  0 & 0 & 0 & 0 & 0 & 0 & 0 & 0 & 0\\
  0 & -p & 0 & -p & 0 & 0 & 0 & 0 & 0 \\
  -p & 0 & 0 & 0 & p & 0 & 0 & 0 & 0 \\
  0 & 0& 0& 0& 0& 0& 0& 0& 0 \\
  0& 0& 0& 0& 0& 0& 0& 0& 0 \\
  0& 0& 0& 0& 0& 0& 0& 0& 0 \\
  0& 0& 0& 0& 0& 0& 0& 0& p 
 
\end{pmatrix}
.
\end{equation}
The singular values ($ \lambda_{i} $) of the matrix M are $ \lbrace 2p,\, 2p,\, p \rbrace $. Consequently, the upper bound of the maximal value of the SS operators is \\
\begin{align*}
 V(S_{4}) = \max \vert \langle S_{4}\rangle _{\rho} \vert_{\theta  = \frac{\pi}{4}} & = 4 \sqrt{2} .2p, \\
& = 8 \sqrt{2}p.
 \end{align*}
Here the singular vectors corresponding to $ \lambda_{1,2} $ can be selected as $ (1,0,0,0,-1,0,0,0,0)^{T} $ and $ (0,1,0,1,0,0,0,0,0)^{T} $, which can be decomposed as follow,\\
$ (1,0,0,0,-1,0,0,0,0)^{T} = (1,0,0)^{T} \otimes (1,0,0)^{T}- (0,1,0)^{T} \otimes (0,1,0)^{T} $,\\
and $ (0,1,0,1,0,0,0,0,0)^{T} = (1,0,0)^{T} \otimes (0,1,0)^{T}+ (0,1,0)^{T} \otimes (1,0,0)^{T} $.\\ Hence, we can set $ \overrightarrow{a} = (1,0,0) $, $ \overrightarrow{a'} = (0,1,0) $ and $ \overrightarrow{b} = (1,0,0) $, $ \overrightarrow{b'} = (0,1,0)  $. Considering the above settings, one can find that each of the inequalities in the proof of the Theorem 1 becomes equal, which means that the upper bound is saturated for $ \rho_{\theta = \frac{\pi}{4}} $. \\
However, we have observed that whenever we set $ p = 1 $ in \eqref{b5} the optimal value of the four part SS operator is 
\begin{align*}\label{b6}
V(S_{4}) & = 4\sqrt{2}.2\sin2\theta\\
& = 8 \sqrt{2} \sin 2\theta. 
\end{align*}
Clearly, our result is in accordance with the main result of SS nonlocality in Ref.\cite{19}. We have observed that the mixed state $ \rho  $ in Eq. \eqref{b5} violates the SS inequality and exhibit genuine four qubit nonlocality if $ p > \frac{\sqrt{2}}{2 \sin 2 \theta}.$ We have numerically evaluated the threshold value of $ \theta $ for $ \rho $ in Eq. \eqref{b5} to exhibit genuine four qubit  nonlocality, which is $ \theta > \frac{\pi}{8} $ and $ 0.7071<p \leq 1 $ and our theorem is consistent with this result.\\
\textbf{Example 2.} We have considered the mixed state $ \sigma $ consist of maximal slice (MS) state \cite{46} with noise.
\begin{equation}\label{b7}
\sigma = p \vert \phi\rangle \langle \phi\vert + \frac{1-p}{8} I_{2}^{\otimes 4},
\end{equation}
where $ \vert \phi \rangle = \frac{1}{\sqrt{2}} ( \vert 0000\rangle + \vert 111\rangle (\cos \theta \vert 0\rangle + \sin \theta \vert 1 \rangle )) $ is the four qubit MS state \cite{44}. From our Theorem 1, the singular values of the correlation matrix of the state $ \sigma $ in \eqref{b7} are $ \lbrace p \sin \theta ,\,p \sqrt{2(1+\sin^{2}\theta)},\,p \sqrt{2 (1+\sin^{2}\theta)} \rbrace  $. Hence the maximum bound of the four part SS operator is  
\begin{align*}
V(S_{4}) & = 4\sqrt{2}.p \sqrt{2(1+\sin^{2}\theta)}\\
& = 8p \sqrt{(1+\sin^{2}\theta)}.
\end{align*}
Numerical optimization suggest that our result is in accordance with the the bounds of SS operator in Ref. \cite{19}. Consequently $ \sigma $ will demonstrate genuine four qubit nonlocality if $ p> \frac{1}{\sqrt{(1+\sin^{2}\theta)}} $.\\

\subsubsection{Tight upper bound for n-qubit system} 
The Seevinck and Svetlichny inequality for n-qubit system is given by \cite{19}\\
\begin{equation}
\vert \langle S^{\pm}_{n} \rangle \vert \leq 2^{n-1},
\end{equation}
where $  S^{ \pm }_{n} = \sum \limits_{I} \nu^{ \pm }_{t(I)} A^{1}_{i_{1}}..... A^{n}_{i_{n}}, $ $ I=( i_{1}, i_{2},.....i_{n}) (i=1,2) $  and $ t(I) $ is the number of times index 2 appears in $ I $. The sequence of signs $ \nu^{ \pm }_{k} $ is given by:
\begin{equation}
\nu^{ \pm }_{k} = (-1)^{k(k\pm 1)/2}. 
 \end{equation}
 \textbf{Hypothesis:} For an arbitrary n-qubit state $ \rho $ the optimal quantum bound of the Seevinck and Svetlichny operator is $ V(S_{n})= (\sqrt{2})^{n+1} \lambda_{\max }, $ where $  \lambda_{\max }  $ is the largest singular value of the correlation matrix $ M $.\\
 The real coefficients of the matrix $ M $ are given by,\\ 
$  M = [t_{j_{1}j_{2}....j_{n}}],$ with $ t_{j_{1}j_{2}....j_{n}}= Tr[\rho(\sigma_{j_{1}}\otimes\sigma_{j_{2}}\otimes.....\sigma_{j_{n}})] $,
where $ \sigma_{j_{n}} $ are  the Pauli operators with three orthogonal directions $ j_{n} = 1,2,3. $\\
The above hypothesis is based on the results of 5-qubit and 6-qubit Seevinck and Svetlichny operator ( see Appendix \eqref{a3} and \eqref{a4}).

\section{Four party genuine entanglement and Nonlocality of Pure state}
For quantum states with more than two qubits there is no single measure of entanglement, and therefore no unique maximally entangled state \cite{47}. Meyer and Wallach \cite{48} defined a single parameter measure of pure state entanglement for three and four qubit states. This measure was further explored by Brennen \cite{49} who showed that it is a monotone. For the pure state $ \vert \psi \rangle $, the Meyer-Wallach (MW) measure written in the Brennen form is:
\begin{equation}
 S(\vert \psi \rangle) =\frac{1}{n} \sum_{k=1}^{n} 2 (1- Tr[\rho_{k}^{2}])
 \end{equation} 
 where $ \rho_{k} $ is the one-qubit reduced density matrix of the $ k$th qubit after tracing out the rest.\\
 The MW measure was originally described as a measure of global entanglement, to distinguish it from purely bipartite measures such as the concurrence. However, it is not able to distinguish states which are fully inseparable from states which, while entangled, are separable into states of some set of subsystems. This property becomes a serious drawback in the context of the analysis of experimental data. For example, the MW measure is one both for the four qubit GHZ state and a product of two two qubit Bell states. Later on, Love et. al., \cite{50} defined a global measure of entanglement, this measure is zero except on genuine entangled (fully inseparable) states. this measure for four qubit state is written as:
 \begin{equation}\label{m1}
 \begin{split}
    C_{1234}(\rho) = (C_{1(234)}C_{2(134)}C_{3(124)}C_{4(123)}C_{(12)(34)}*\\
    C_{(13)(24)}C_{(14)(23)})^{\frac{1}{7}}, 
    \end{split}
 \end{equation} 
 where $ C_{A(BCD)} $ and $ C_{AB(CD)} $ are defined as $ C_{A(BCD)} = \sqrt{2(1-Tr(\rho_{A}^{2}))} $ and $ C_{(AB)(CD)} = \sqrt{\frac{4}{3} (1-Tr(\rho_{AB}^{2}))}  $, respectively (where $ 1, 2, 3$ and 4  denotes the parties $ A, B, C $ and D respectively). \\
 In Ref. \cite{29}, the authors have classified four qubit pure states as fully separable, tri-separable, bi-separable and fully inseparable form. This classification is based on the use of generalized Schmidt decomposition of pure states of multiqubit systems \cite{51,52}. While among these classifications, only fully inseparable class of states have non zero concurrence of the form Eq. \eqref{m1}, i.e. $ C_{1234} (\rho) \neq 0 $ for all classes of states which are fully inseparable. It is shown that there is eleven different class of fully inseparable states. Here we have developed a relationship between the entangled measure and our quantity $ V(S_{4}) $ for some class of states ( we have used the same notations as used in Ref. \cite{29}).\\
 (i) $ \vert \psi_{1}\rangle = \alpha \vert 0000\rangle + \omega \vert 1111\rangle$.\\
The singular values of this state are $ \lbrace 1, 4\alpha \omega,  4\alpha \omega \rbrace $, while the  global entanglement $ C_{1234} = 2 \alpha \omega $. Since the largest singular value is degenerate, consequently the bound  $ V(S_{4}) $  is tight, therefore, \\
\begin{align}
V(S_{4}) = 8\sqrt{2}C_{1234},
\end{align}
 $ C_{1234}> \frac{1}{\sqrt{2}}  $ indicates the presence of genuine nonlocality in the said state. Thus global entanglement measure can reveal the presence of genuine nonlocality for this class of states.\\
(ii) $ \vert \psi_{2}\rangle = \alpha \vert 0000\rangle + \kappa \vert 1011\rangle + \mu \vert 1101\rangle $.\\
For this class of states the degenerate set of  largest singular value is $ \max \lbrace 2\sqrt{2} \alpha \kappa, 2\sqrt{2} \alpha \mu \rbrace $, while the global entanglement measure $ C_{1234} = 2[ \frac{\sqrt{2}}{3\sqrt{3}} (\alpha \sqrt{1-\alpha^{2}})^{3} \kappa \mu \sqrt{1-\kappa^{2}} \sqrt{1-\mu^{2}} (1-(\alpha^{4}+\kappa^{4}+\mu^{4}))^{2}]^{\frac{1}{7}}$. However the bound $ V(S_{4})  $ is given by \\
\begin{align}
V(S_{4}) = \max \lbrace 16 \alpha \kappa, 16 \alpha \mu \rbrace,
\end{align}
genuine nonlocality is detected if $ \alpha \kappa > \frac{1}{2} $ or $ \alpha \mu > \frac{1}{2} $.\\
(iii) The class of states $ \vert \psi_{3} \rangle = \alpha \vert 0000\rangle  + \kappa \vert 1011\rangle  + \lambda \vert 1100\rangle  + \mu \vert 1101\rangle + \omega \vert 1111\rangle  $ have degenerate largest singular values, consequently the genuine nonlocality of this class of states can be revealed through the quantity $ V(S_{4}) $.\\
The residue class of states lacks any degenerate largest singular value, consequently, the genuine nonlocality of these classes of state can't be revealed using the operator $ V(S_{4}) $.\\  
\begin{figure}[h]
\includegraphics[scale=0.5]{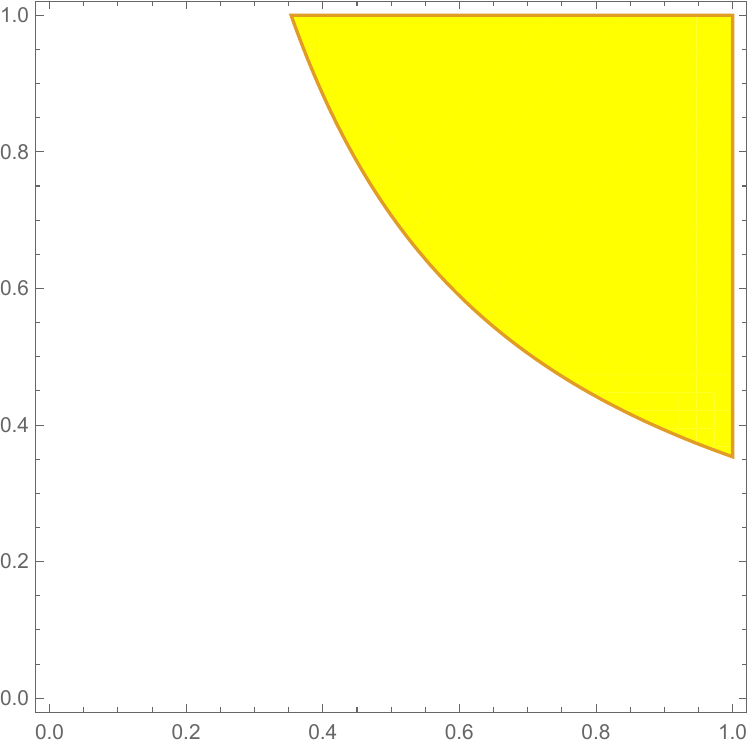}
\caption{Genuine nonlocality of the GHZ class states (i), $ 2\alpha \omega >\frac{1}{\sqrt{2}} $ (yellow region), while it is genuine entangled throughout the region, $ 2\alpha \omega >0,$ where $ \alpha^{2}+\omega^{2}=1. $}
\label{class1}
\end{figure}

\section{Genuine hidden nonlocality under local filtering}
\begin{figure}[h]
\includegraphics[scale=0.5]{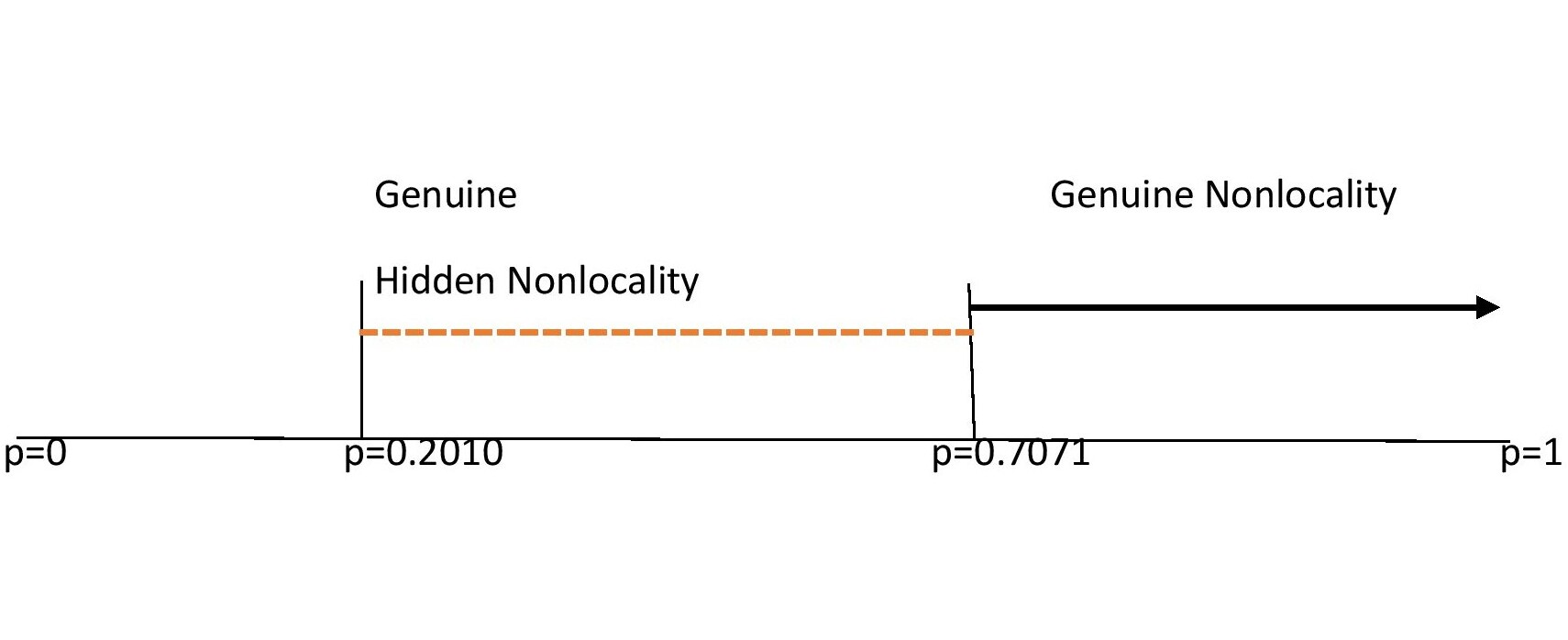}
\caption{Genuine hidden nonlocality under local filtering. Initially, the state $ \rho $ exhibit genuine nonlocality for $ p > .7071 $ and after local filtering it shows genuine nonlocality for $ p \geq .2010 $. Hence it reveal genuine hidden nonlocality in the range $ [.2010,0.7071]. $}
\label{upper1}
\end{figure}

For any four qubit state $ \rho $, under local filtering operation, one gets \cite{53},
\begin{equation}\label{ll1}
\rho' = \frac{1}{N'} (F_{A} \otimes F_{B} \otimes F_{C} \otimes F_{D}) \rho (F_{A} \otimes F_{B} \otimes F_{C} \otimes F_{D})^{\dagger},
\end{equation}
where $ N' = Tr[ (F_{A} \otimes F_{B} \otimes F_{C} \otimes F_{D}) \rho (F_{A} \otimes F_{B} \otimes F_{C} \otimes F_{D})^{\dagger}]$ is the normalization factor, while the filter operators $ F_{X}\; ( X = A,\,B,\,C,\,D) $ are acting on the local subsystems $ X $ accordingly. Let, $ F_{A} = K \Sigma_{A} K^{\dagger} $, $ F_{B} = L \Sigma_{A} L^{\dagger} $, $ F_{C} = M \Sigma_{A} M^{\dagger} $ and $ F_{D} = N\Sigma_{D} N^{\dagger} $ be the spectral decomposition of the filter operators $ F_{A} $, $ F_{B} $, $ F_{C} $ and $ F_{D} $ respectively, where $ K $, $ L $, $ M $ and $ N $ are the unitary operators. Set, $ \alpha_{p} = \Sigma_{A} \sigma_{p} \Sigma_{A} $, $ \beta_{q} = \Sigma_{B} \sigma_{q} \Sigma_{B} $, $ \gamma_{r} = \Sigma_{C} \sigma_{r} \Sigma_{C} $, and $ \delta_{s} = \Sigma_{D} \sigma_{s} \Sigma_{D} $. Without loss of generality, we assume that
\begin{equation}\label{ll2}
\Sigma_{A}=
\begin{pmatrix}
x & 0\\
0 & 1
\end{pmatrix}
, \Sigma_{B} =
\begin{pmatrix}
y & 0\\
0 & 1
\end{pmatrix}
, \Sigma_{C} =
\begin{pmatrix}
z & 0\\
0 & 1
\end{pmatrix}
and\; \Sigma_{D} =
\begin{pmatrix}
t & 0\\
0 & 1
\end{pmatrix}
,
\end{equation}
where $ x,\,y,\, z,\,t \geq 0 $. \\
Consider, $ \Lambda = [\Lambda_{rs,pq}] $ be a matrix whose elements are given by,
\begin{equation}\label{lr3}
\Lambda_{pqrs} = Tr[ \sigma(\alpha_{p} \otimes \beta_{q} \otimes \gamma_{r} \otimes \delta_{s})],\;\; p,\,q,\,r,\,s\, = 1,\, 2,\, 3,
\end{equation}
where $ \sigma $ is any state that is locally unitary equivalent to $ \rho $.\\
\textbf{Theorem 2.} For any four qubit locally filtered state $ \rho' = \frac{1}{N'} (F_{A} \otimes F_{B} \otimes F_{C} \otimes F_{D}) \rho (F_{A} \otimes F_{B} \otimes F_{C} \otimes F_{D})^{\dagger} $ of $ \rho $ the optimal quantum bound of SS operator in Eq. \eqref{b1} is given by,
\begin{equation}\label{ll4}
V(S_{4})' = \max \vert \langle S_{4}\rangle_{\rho'} \vert\leq 4 \sqrt{2} \lambda_{\max}^{'},
\end{equation}
where $\langle S_{4}\rangle_{\rho'} = Tr(S_{4} \rho')$, and $ \lambda'_{\max} $ is the maximal singular value of the matrix $ \frac{\Lambda}{N'} $, where $ \Lambda $ is defined in Eq. \eqref{lr3} taking over all quantum states, that are locally unitary equivalent to $ \rho $. Consequently, $ \lambda'_{\max} $  is also the maximal singular value of the matrix $ M' = [t'_{ijkl}] $, where $ t'_{ijkl} = Tr [ \rho' (\sigma_{i} \otimes \sigma_{j} \otimes \sigma_{k} \otimes \sigma_{l})],\; i,j,k,l = 1,2,3$ (see appendix \eqref{aa1} for proof).\\
 We have considered a four qubit entangled state, and by applying local filtering operation, it reveals hidden nonlocality of that state in this scenario.\\
\textbf{Example 3.} Considered the state,
\begin{equation}\label{ll8}
\rho = p \vert \psi_{\theta} \rangle \langle \psi_{\theta} \vert +\frac{1-p}{4} \vert 00\rangle \langle 00\vert \otimes I_{2}^{\otimes 2},
\end{equation}
where $  \vert \psi_{\theta} \rangle =  ( \cos \theta \vert 0000\rangle + \sin \theta \vert 1111\rangle), $  $ 0\leq p\leq 1, $ $ 0<\theta\leq \frac{\pi}{4}, $ and $ I_{2} $ is the identity matrix. The correlation matrix $ G $ is of the following form,\\
\begin{equation}
G = 
\begin{pmatrix}

  k & 0 & 0 & 0 & -k & 0 & 0 & 0 & 0\\
 0 & -k & 0 & -k & 0 & 0 & 0 & 0 & 0\\
  0 & 0 & 0 & 0 & 0 & 0 & 0 & 0 & 0\\
  0 & -k & 0 & -k & 0 & 0 & 0 & 0 & 0 \\
  -k & 0 & 0 & 0 & k & 0 & 0 & 0 & 0 \\
  0 & 0& 0& 0& 0& 0& 0& 0& 0 \\
  0& 0& 0& 0& 0& 0& 0& 0& 0 \\
  0& 0& 0& 0& 0& 0& 0& 0& 0 \\
  0& 0& 0& 0& 0& 0& 0& 0& p 
 
\end{pmatrix}
,
\end{equation}
where $ k = p \sin 2\theta.$ The singular values $ (\lambda) $ of the matrix $ G $ are $ \lbrace 2p\sin 2\theta,\, 2p \sin 2\theta,\, p \rbrace $. Consequently, the upper bound of the maximal value of the SS operators is \\
\begin{align*}
 V(S_{4}) = \max \vert \langle S_{4}\rangle _{\rho} \vert & = 4 \sqrt{2} .2p \sin 2\theta, \\
& = 8 \sqrt{2}p \sin 2\theta.
 \end{align*}
 The singular vectors corresponding to $ \lambda_{\max} $ can be selected as $ (1,0,0,0,-1,0,0,0,0)^{T} $ and $ (0,1,0,1,0,0,0,0,0)^{T} $, which can be decomposed as follow,\\
$ (1,0,0,0,-1,0,0,0,0)^{T} = (1,0,0)^{T} \otimes (1,0,0)^{T}- (0,1,0)^{T} \otimes (0,1,0)^{T} $,\\
and $ (0,1,0,1,0,0,0,0,0)^{T} = (1,0,0)^{T} \otimes (0,1,0)^{T}+ (0,1,0)^{T} \otimes (1,0,0)^{T} $.\\ Hence, we can set $ \overrightarrow{a} = (1,0,0) $, $ \overrightarrow{a'} = (0,1,0) $ and $ \overrightarrow{b} = (1,0,0) $, $ \overrightarrow{b'} = (0,1,0)  $. Hence the upper bound is saturated for $ \rho $. \\
We have observed that the state $ \rho_{\theta =\frac{\pi}{4}} $ of Eq. \eqref{ll8} violates the SSI if $ p> 0.7071 $. Hence the state fails to exhibit genuine four particle nonlocality whenever $ 0< p <0.7071 $, here we have shown that the state can reveal hidden nonlocality in the above quoted range. Let us set $ \theta = \frac{\pi}{8} $, it is already shown that for $ \theta = \frac{\pi}{8} $ the state produce optimal value of SS operator is $ V(S_{4}) = 8p $, it clearly respect the SSI \eqref{b2}.\\
Now we apply local filtering on $ \rho $. The correlation matrix of $ \rho $ after the local filtering is given by, 
\begin{equation}
G^{'} = 
\begin{pmatrix}

  k_{1} & 0 & 0 & 0 & -k_{1} & 0 & 0 & 0 & 0\\
 0 & -k_{1} & 0 & -k_{1} & 0 & 0 & 0 & 0 & 0\\
  0 & 0 & 0 & 0 & 0 & 0 & 0 & 0 & 0\\
  0 & -k_{1} & 0 & -k_{1} & 0 & 0 & 0 & 0 & 0 \\
  -k_{1} & 0 & 0 & 0 & k_{1} & 0 & 0 & 0 & 0 \\
  0 & 0& 0& 0& 0& 0& 0& 0& 0 \\
  0& 0& 0& 0& 0& 0& 0& 0& 0 \\
  0& 0& 0& 0& 0& 0& 0& 0& 0 \\
  0& 0& 0& 0& 0& 0& 0& 0& d_{1} 
 
\end{pmatrix}
.
\end{equation}
where $ k_{1}= ptxyz \sin 2\theta $, $ d_{1} = \frac{1}{4} (1 - p) x^2 y^2 - \frac{1}{4} (1 - p) t^2 x^2 y^2 -  \frac{1}{4} (1 - p) x^2 y^2 z^2 +  t^2 x^2 y^2 z^2 (\frac{1}{4} + p (-\frac{1}{4} + \cos^{2} \theta)) +  p \sin^{2} \theta $ and the normalization factor $ N' = \frac{1}{4} (1 - p) x^2 y^2 + \frac{1}{4} (1 - p) t^2 x^2 y^2 +  \frac{1}{4} (1 - p) x^2 y^2 z^2 +  t^2 x^2 y^2 z^2 (\frac{1}{4} + p (-\frac{1}{4} + \cos^{2} \theta)) +  p \sin^{2} \theta $. The singular values of the matrix $ G^{'} $ are $ 2ptxyz \sin 2\theta,\,2ptxyz \sin 2\theta $ and $ d_{1} $. Since locally unitary equivalence states $ \rho $ and $ \varrho $ have the same set of singular values, hence $ \frac{2ptxyz \sin 2 \theta}{N'},\,\frac{2ptxyz\sin 2 \theta}{N'}  $ and $ \frac{d_{1}}{N'} $ are the singular values of the matrix $ \frac{\Lambda}{N'} $, consequently these are the singular values of the matrix $ M' $ of Theorem 2. The maximal singular value $ \lambda'_{\max} $ of $ M' $ is $  \frac{2ptxyz\sin 2\theta}{N'} $, provided $  \frac{2ptxyz\sin 2\theta}{N'}> \frac{d_{1}}{N'} $. Therefore the upper bound of the maximal value of SS operator is given by, 
\begin{equation}
V(S_{4})^{'} = \max \vert \langle S_{4}\rangle_{\rho'} \vert \leq \frac{8 \sqrt{2} pxyzt \sin 2\theta}{N'}.
\end{equation} 
Since there is degeneracy in $ \lambda'_{\max} $, the matrix $ \frac{\Lambda}{N'} $ has the singular vectors $ \overrightarrow{v_{1}} $ and  $ \overrightarrow{v_{2}} $ with respect to the singular value $ \lambda'_{\max} $. In accordance with Theorem 2, the singular vectors of $ M' $ corresponding to $ \lambda'_{\max} $  are  $ (O_{C} \otimes O_{D})\overrightarrow{v_{1}} = O_{C} \overrightarrow{a} \otimes O_{D} \overrightarrow{b} -  O_{C} \overrightarrow{a'} \otimes O_{D} \overrightarrow{b'} $ and $ (O_{C} \otimes O_{D})\overrightarrow{v_{2}} = O_{C} \overrightarrow{a} \otimes O_{D} \overrightarrow{b'} +  O_{C} \overrightarrow{a'} \otimes O_{D} \overrightarrow{b} $. Hence the state violates the SS inequality if $  \frac{ptxyz \sin 2\theta}{N'} > \frac{\sqrt{2}}{2} $, with the constrain $ \frac{2ptxyz \sin 2\theta}{N'}> \frac{d_{1}}{N'} $. Using numerical optimization, we have observed that for nonzero $ \theta $, the state $ \rho $ violates the SS inequality and exhibit genuine four qubit nonlocality for $ 0.2010 \leq p \leq 1 $, thus this state exhibit genuine hidden nonlocality in the range $ 0.2010 \leq p\leq 0.7071 $ (see Fig. \ref{upper1}).\\

 \section{Conclusion}
In this work, we have presented a quantitative analysis of the genuine four party nonlocality for some four qubit quantum systems via effective computation of the maximal quantum value of the SS operators. Our method provides a tight upper bound of the maximal quantum value of SS operator. The tightness of the bound is investigated through several noisy quantum states. Our results provide an effective and operational method to detect genuine four party nonlocality. Consequently, a relationship between nonlocality and four party entanglement, i.e., concurrence of some class of states have been discussed, and we have provided a lower bound on concurrence for GHZ class of states to exhibit genuine nonlocality. Further We have presented a qualitative analysis of the hidden genuine nonlocality for four qubit systems by providing a tight upper bound on the maximal quantum value of the SS operators under local filtering operations. We have presented a class of four qubit states whose hidden genuine nonlocality can be revealed by local filtering. One may find this results helpful in investigating trade off relations \citep{56} in genuine four qubit nonlocality. The methods presented in this paper can also be used in computing the maximal violations of other four party or multipartite Bell-type inequalities such as the Refs. \cite{23,24,57}. We hope our study will be helpful for a better understanding of multipartite nonlocality as well as for executing future quantum protocols based on multipartite nonlocality.  
 \section{Acknowledgements}
 The authors like to thank A. Halder for his suggestion and the authors IC and Debasis Sarkar acknowledges the work as part of Quest initiatives by DST India. 
 \appendix
 \section{Proof of Lemma. 1}\label{a1}
 Lemma. Let A be a rectangular matrix of size $ m \times n $. For any vector $ \overrightarrow{x} \in R^{m} $ and $ \overrightarrow{y} \in R^{n} $, we have\\
\begin{equation}\label{b3}
  \vert \overrightarrow{x}^{T} A \overrightarrow{y} \vert \leq \lambda_{\max}\vert\overrightarrow{x}\vert \vert\overrightarrow{y}\vert,
    \end{equation}  
    where $ \lambda_{\max} $ is the largest singular value of matrix $A$. The bound is tight when $ \overrightarrow{x} $ and $ \overrightarrow{y} $ are the corresponding singular vectors of $ A $ concerning $ \lambda_{\max} $ .
 \begin{proof}
 By the singular value decomposition, there exist two unitary matrices $U$ and $V$ such that $ A = U^{T} \Sigma V $, where $ \Sigma $ has only nonzero elements along with its diagonal. Therefore, we may
assume that $ A =\Sigma $ and consider only the following form, $ G(\overrightarrow{x},\overrightarrow{y}) = \sum\limits_{i} a_{i}x_{i}y_{i} $, where $ a_{1} \geq a_{2 }\geq.....\geq a_{n} $. Using
the Cauchy-Schwarz inequality for the inner product $ \langle\overrightarrow{x} \vert \overrightarrow{y}\rangle := G(\overrightarrow{x},\overrightarrow{y}) $, we have that

\begin{align*}
 |G(\overrightarrow{x},\overrightarrow{y})|& \leq G(\overrightarrow{x},\overrightarrow{x})^{\frac{1}{2}}  G(\overrightarrow{y},\overrightarrow{y})^{\frac{1}{2}}\\
&  = (  \sum\limits_{i} a_{i}x^{2}_{i})^{\frac{1}{2}} (\sum\limits_{i} a_{i}y^{2}_{i})^{\frac{1}{2}}   \\
&  \leq a_{1} (\sum\limits_{i} x^{2}_{i})^{\frac{1}{2}} (\sum\limits_{i} y^{2}_{i})^{\frac{1}{2}}
\end{align*}
Here $ a_{1} $ signifies $ \lambda_{\max} $ in Eq. \eqref{b3}.  
 \end{proof}
 \section{Proof of Theorem. 1 }\label{a2}
Theorem 1. For any four qubit quantum state $ \rho $, the maximum quantum value $ V(S_{4})$ of the SS operator $ S_{4}$ defined in \eqref{b1} is bounded by\\
\begin{equation}\label{k1}
 V(S_{4})= \max \vert \langle S_{4}\rangle_{\rho}\vert \leq 4\sqrt{2} \lambda_{\max}.
\end{equation}
Where $ \langle S_{4}\rangle_{\rho} = $ Tr$(S_{4}\rho)  $, $ \lambda_{\max} $ is the largest singular value of the correlation matrix $ M = [t_{kl,ij}] $, where the elements are given by, $ t_{ijkl} =  $ Tr[$ \rho (\sigma_{i} \otimes \sigma_{j} \otimes \sigma_{k} \otimes \sigma_{l}) $],  $ i,j,k,l =1,2,3 $. \\
 \begin{proof}
By the definition of four particle genuine non separability,\\            
$ S_{4} = [A \otimes B-A' \otimes B'] \otimes [(C-C') \otimes D-(C+C') \otimes D']-[A' \otimes B + A \otimes B'] \otimes [(C+C') \otimes D+(C-C') \otimes D']. $\\
 $  = \sum\limits_{i,j,k,l} [(a_{i}b_{j}-a'_{i}b'_{j})[(c_{k}-c'_{k})d_{l}-(c_{k}+c'_{k})d'_{l}]-(a'_{i}b_{j}+a_{i}b'_{j})[(c_{k}+c'_{k})d_{l}+(c_{k}-c'_{k})d'_{l}] ]\sigma_{i} \otimes \sigma_{j} \otimes \sigma_{k} \otimes \sigma_{l}$\\
 Hence, $ \langle S_{4}\rangle =  \sum\limits_{i,j,k,l} [(a_{i}b_{j}-a'_{i}b'_{j})[(c_{k}-c'_{k})d_{l}-(c_{k}+c'_{k})d'_{l}]-(a'_{i}b_{j}+a_{i}b'_{j})[(c_{k}+c'_{k})d_{l}+(c_{k}-c'_{k})d'_{l}] ]\,Tr[\rho(\sigma_{i} \otimes \sigma_{j} \otimes \sigma_{k} \otimes \sigma_{l})] $\\
 $  =  \sum\limits_{i,j,k,l} [(a_{i}b_{j}-a'_{i}b'_{j})[(c_{k}-c'_{k})d_{l}-(c_{k}+c'_{k})d'_{l}]-(a'_{i}b_{j}+a_{i}b'_{j})[(c_{k}+c'_{k})d_{l}+(c_{k}-c'_{k})d'_{l}] ]\,t_{ijkl} $\\
 $ = (\overrightarrow{a} \otimes \overrightarrow{b}-\overrightarrow{a'} \otimes \overrightarrow{b'})^{T} M [(\overrightarrow{c}-\overrightarrow{c'}) \otimes \overrightarrow{d}-(\overrightarrow{c}+\overrightarrow{c'}) \otimes \overrightarrow{d'}] - (\overrightarrow{a'} \otimes \overrightarrow{b}+\overrightarrow{a} \otimes \overrightarrow{b'})^{T} M [(\overrightarrow{c}+\overrightarrow{c'}) \otimes \overrightarrow{d}+(\overrightarrow{c}-\overrightarrow{c'}) \otimes \overrightarrow{d'}]$.\\
 Therefore from the above lemma \eqref{b3} we have,\\
 $ \vert \langle S_{4}\rangle\vert \leq \lambda_{\max} [ \vert (\overrightarrow{a} \otimes \overrightarrow{b}-\overrightarrow{a'} \otimes \overrightarrow{b'})\vert\, \vert (\overrightarrow{c}-\overrightarrow{c'}) \otimes \overrightarrow{d}-(\overrightarrow{c}+\overrightarrow{c'}) \otimes \overrightarrow{d'}\vert + \vert (\overrightarrow{a'} \otimes \overrightarrow{b}+\overrightarrow{a} \otimes \overrightarrow{b'})\vert\, \vert (\overrightarrow{c}+\overrightarrow{c'}) \otimes \overrightarrow{d}+(\overrightarrow{c}-\overrightarrow{c'}) \otimes \overrightarrow{d'}\vert ]$.\\
 Let, $ \theta_{a} $, be the angle between $ \overrightarrow{a} $ and $ \overrightarrow{a'} $, accordingly we define $ \theta_{b}\, $, $ \theta_{c}\, $ and $ \theta_{d}\, $ among the measurement directions of the parties B, C and D.\\
 We have, $ \vert (\overrightarrow{a} \otimes \overrightarrow{b}-\overrightarrow{a'} \otimes \overrightarrow{b'})\vert^{2} = 2-2 \langle \overrightarrow{a},\overrightarrow{a'}\rangle \langle \overrightarrow{b},\overrightarrow{b'}\rangle = 2-2 \cos \theta_{a} \cos \theta_{b} $,\\
 and  $ \vert (\overrightarrow{a'} \otimes \overrightarrow{b}+\overrightarrow{a} \otimes \overrightarrow{b'})\vert^{2} = 2+2 \langle \overrightarrow{a},\overrightarrow{a'}\rangle \langle \overrightarrow{b},\overrightarrow{b'}\rangle = 2+2 \cos \theta_{a} \cos \theta_{b} $.\\
 Let us consider the principal angle $\theta_{ab}  $ such that, $ \cos \theta_{a} \cos \theta_{b} = \cos \theta_{ab} $.\\
 Consequently,\\
  $  \vert (\overrightarrow{a} \otimes \overrightarrow{b}-\overrightarrow{a'} \otimes \overrightarrow{b'})\vert^{2} = 4 \sin^{2} \frac{\theta_{ab}}{2} $,\\
   $  \vert (\overrightarrow{a'} \otimes \overrightarrow{b}+\overrightarrow{a} \otimes \overrightarrow{b'})\vert^{2} = 4 \cos^{2} \frac{\theta_{ab}}{2}  $, \\
 $ \vert (\overrightarrow{c}-\overrightarrow{c'}) \otimes \overrightarrow{d}-(\overrightarrow{c}+\overrightarrow{c'}) \otimes \overrightarrow{d'}\vert^{2} = 4, $\\
 $  \vert (\overrightarrow{c}+\overrightarrow{c'}) \otimes \overrightarrow{d}+(\overrightarrow{c}-\overrightarrow{c'}) \otimes \overrightarrow{d'}\vert^{2} = 4 $ .\\ 
 Therefore,
   \begin{align*}
   \vert \langle S_{4}\rangle \vert  & \leq   4 \lambda_{1} [\vert \sin\frac{\theta_{ab}}{2}\vert + \vert \cos \frac{\theta_{ab}}{2}\vert ] \\
  & \leq 4 \sqrt{2} \lambda_{\max} 
\end{align*}   
 Hence the theorem.
 
\end{proof} 
\section{Tight upper bound of 5-qubit Seevinck and Svetlichny operator }\label{a3}
\begin{proof}
The 5-qubit Seevinck and Svetlichny operator is \citep{19},\\
  $ S_{5} = (A \otimes B'+A' \otimes B)\otimes [(C	\otimes D-C' \otimes D')\otimes (E-E')- (C	\otimes D'+C' \otimes D)\otimes (E+E')] + (A \otimes B-A' \otimes B')\otimes [(C	\otimes D'+C' \otimes D)\otimes (E-E')+ (C	\otimes D-C' \otimes D')\otimes (E+E')]$.\\
  =$ \sum \limits_{i,j,k,l,m} [(a_{i}b'_{j}+a'_{i}b_{j})[(c_{k}d_{l}-c'_{k}d'_{l})(e_{m}-e'_{m})- (c_{k}d'_{l}+c'_{k}d_{l})(e_{m}+e'_{m})]+ (a_{i}b_{j}-a'_{i}b'_{j})[(c_{k}d'_{l}+c'_{k}d_{l})(e_{m}-e'_{m})+ (c_{k}d_{l}-c'_{k}d'_{l})(e_{m}+e'_{m})]] \sigma_{i} \otimes \sigma_{j} \otimes \sigma_{k} \otimes \sigma_{l} \otimes \sigma_{m} $.\\
 Hence,  $ \langle S_{5} \rangle = \sum \limits_{i,j,k,l,m} [(a_{i}b'_{j}+a'_{i}b_{j})[(c_{k}d_{l}-c'_{k}d'_{l})(e_{m}-e'_{m})- (c_{k}d'_{l}+c'_{k}d_{l})(e_{m}+e'_{m})]+ (a_{i}b_{j}-a'_{i}b'_{j})[(c_{k}d'_{l}+c'_{k}d_{l})(e_{m}-e'_{m})+ (c_{k}d_{l}-c'_{k}d'_{l})(e_{m}+e'_{m})]] Tr[ \rho (\sigma_{i} \otimes \sigma_{j} \otimes \sigma_{k} \otimes \sigma_{l} \otimes \sigma_{m})] $\\
 =$ \sum \limits_{i,j,k,l,m} [(a_{i}b'_{j}+a'_{i}b_{j})[(c_{k}d_{l}-c'_{k}d'_{l})(e_{m}-e'_{m})- (c_{k}d'_{l}+c'_{k}d_{l})(e_{m}+e'_{m})]+ (a_{i}b_{j}-a'_{i}b'_{j})[(c_{k}d'_{l}+c'_{k}d_{l})(e_{m}-e'_{m})+ (c_{k}d_{l}-c'_{k}d'_{l})(e_{m}+e'_{m})]] t_{ijklm} $,\\
$ \Rightarrow \langle S_{5} \rangle  =  (\overrightarrow{a} \otimes \overrightarrow{b'} + \overrightarrow{a'} \otimes \overrightarrow{b})^{T} M [(\overrightarrow{c} \otimes \overrightarrow{d} -\overrightarrow{c'} \otimes \overrightarrow{d'}) \otimes (\overrightarrow{e}-\overrightarrow{e'})- (\overrightarrow{c} \otimes \overrightarrow{d'}+\overrightarrow{c'} \otimes \overrightarrow{d}) \otimes (\overrightarrow{e}+\overrightarrow{e'})]+ (\overrightarrow{a} \otimes \overrightarrow{b} - \overrightarrow{a'} \otimes \overrightarrow{b'})^{T} M [(\overrightarrow{c} \otimes \overrightarrow{d'}+\overrightarrow{c'} \otimes \overrightarrow{d}) \otimes (\overrightarrow{e}-\overrightarrow{e'})+ (\overrightarrow{c} \otimes \overrightarrow{d}-\overrightarrow{c'} \otimes \overrightarrow{d'}) \otimes (\overrightarrow{e}+\overrightarrow{e'})]$.\\
 Therefore from the above lemma \eqref{b3} we have,\\
$ \vert \Rightarrow \langle S_{5} \rangle \vert \leq \lambda_{\max }[ \vert (\overrightarrow{a} \otimes \overrightarrow{b'} + \overrightarrow{a'} \otimes \overrightarrow{b}) \vert \; \vert (\overrightarrow{c} \otimes \overrightarrow{d}-\overrightarrow{c'} \otimes \overrightarrow{d'}) \otimes (\overrightarrow{e}-\overrightarrow{e'})- (\overrightarrow{c} \otimes \overrightarrow{d'}+\overrightarrow{c'} \otimes \overrightarrow{d}) \otimes (\overrightarrow{e}+\overrightarrow{e'}) \vert + \vert (\overrightarrow{a} \otimes \overrightarrow{b} - \overrightarrow{a'} \otimes \overrightarrow{b'})\vert \; \vert (\overrightarrow{c} \otimes \overrightarrow{d'}+\overrightarrow{c'} \otimes \overrightarrow{d}) \otimes (\overrightarrow{e}-\overrightarrow{e'})+ (\overrightarrow{c} \otimes \overrightarrow{d}-\overrightarrow{c'} \otimes \overrightarrow{d'}) \otimes (\overrightarrow{e}+\overrightarrow{e'})\vert ]$.  \\
Let us consider, $ \theta_{a} $, be the angle between $ \overrightarrow{a} $ and $ \overrightarrow{a'} $, accordingly we define $ \theta_{b}\, $, $ \theta_{c}\, $  $ \theta_{d}\, $ and $ \theta_{e} $ among the measurement directions of the parties B, C, D and E.\\
Considering the techniques used in above Appendix \eqref{a2} we have,\\
$ \vert ( \overrightarrow{a} \otimes \overrightarrow{b} - \overrightarrow{a'} \otimes \overrightarrow{b'}) \vert^{2} = 4 \sin^{2}\frac{ \theta_{ab}}{2}$,\\
 $  \vert (\overrightarrow{a} \otimes \overrightarrow{b'} + \overrightarrow{a'} \otimes \overrightarrow{b}) \vert^{2} = 4 \cos^{2}\frac{\theta_{ab}}{2} $,\\
 $ \vert (\overrightarrow{c} \otimes \overrightarrow{d}-\overrightarrow{c'} \otimes \overrightarrow{d'}) \otimes (\overrightarrow{e}-\overrightarrow{e'})- (\overrightarrow{c} \otimes \overrightarrow{d'}+\overrightarrow{c'} \otimes \overrightarrow{d}) \otimes (\overrightarrow{e}+\overrightarrow{e'}) \vert^{2}=  8 [1+ \cos \theta_{c} \cos \theta_{d} \cos\theta_{e}]$,\\
and $ \vert (\overrightarrow{c} \otimes \overrightarrow{d'}+\overrightarrow{c'} \otimes \overrightarrow{d}) \otimes (\overrightarrow{e}-\overrightarrow{e'})+ (\overrightarrow{c} \otimes \overrightarrow{d}-\overrightarrow{c'} \otimes \overrightarrow{d'}) \otimes (\overrightarrow{e}+\overrightarrow{e'}) \vert^{2}=  8 [1- \cos \theta_{c} \cos \theta_{d} \cos\theta_{e}]$. \\
  Let us consider the principal angle $ \theta_{cde}  $ such that, $ \cos \theta_{c} \cos \theta_{d} \cos \theta_{e} = \cos \theta_{cde} $.\\
Consequently,\\
$   \vert (\overrightarrow{c} \otimes \overrightarrow{d}-\overrightarrow{c'} \otimes \overrightarrow{d'}) \otimes (\overrightarrow{e}-\overrightarrow{e'})- (\overrightarrow{c} \otimes \overrightarrow{d'}+\overrightarrow{c'} \otimes \overrightarrow{d}) \otimes (\overrightarrow{e}+\overrightarrow{e'}) \vert^{2}=  16 \cos^{2} \frac{\theta_{cde}}{2}$, \\
and $ \vert (\overrightarrow{c} \otimes \overrightarrow{d'}+\overrightarrow{c'} \otimes \overrightarrow{d}) \otimes (\overrightarrow{e}-\overrightarrow{e'})+ (\overrightarrow{c} \otimes \overrightarrow{d}-\overrightarrow{c'} \otimes \overrightarrow{d'}) \otimes (\overrightarrow{e}+\overrightarrow{e'}) \vert^{2} = 16 \sin^{2}\frac{\theta_{cde}}{2} $.\\
Therefore,
   \begin{align*}
   \vert \langle S_{5}\rangle \vert  & \leq   8 \lambda_{\max} [\vert \cos (\frac{\theta_{ab}+\theta_{cde}}{2})\vert \\
  & \leq 8 \lambda_{\max}. 
\end{align*}
\end{proof} 
It is clear that the upper bound saturate by proper choice of measurement settings for the parties. Moreover, the resulting inequality saturates if the degeneracy of $ \lambda_{\max}  $ is more than 1, and corresponding to such $ \lambda_{\max}  $ there exist two nine-dimensional singular vectors of the form $ \overrightarrow{a} \otimes \overrightarrow{b'} +\overrightarrow{a'} \otimes \overrightarrow{b} $  and  $ \overrightarrow{a} \otimes \overrightarrow{b} - \overrightarrow{a'} \otimes \overrightarrow{b'} $.\\
As an example, we consider a class of five qubit GHZ state $(\cos \phi \vert 00000\rangle +\sin \phi \vert 11111\rangle ) $. Its optimal quantum violation is given by
\begin{equation}
\langle S_{5} \rangle  =  8 \max \lbrace 2\sqrt{2} \sin 2\phi, 2\sqrt{2} \sin 2\phi, \cos 2\phi \rbrace. 
\end{equation}
  The optimal bound for GHZ state $ ( \phi = \frac{\pi}{4}) $ comes out to be $ 16\sqrt{2} $, maximal for the five qubit SS operator \citep{19}.
 \section{Tight upper bound of 6-qubit Seevinck and Svetlichny operator }\label{a4}
 \begin{proof}
The 6-qubit Seevinck and Svetlichny operator is \citep{19}\\
$ S_{6} = [[(A' \otimes B'-A \otimes B) \otimes (C' \otimes D'-C \otimes D)-(A' \otimes B+ A \otimes B')\otimes (C \otimes D'+C' \otimes D)] \otimes [(E+E') \otimes F+(E-E') \otimes F']+ [(A \otimes B-A' \otimes B') \otimes (C' \otimes D+C \otimes D')+(A' \otimes B+ A \otimes B')\otimes (C \otimes D-C' \otimes D')] \otimes [(E-E') \otimes F-(E+E') \otimes F']$.\\
$ \langle S_{6} \rangle = \sum \limits_{i,j,k,l,m,n} [(a'_{i}b'_{j}-a_{i}b_{j})(c'_{k}d'_{l}-c_{k}d_{l})- (a'_{i}b_{j}+a_{i}b'_{j})(c_{k}d'_{l}+c'_{k}d_{l})][(e_{m}+e'_{m})f_{n}+(e_{m}-e'_{m})f'_{n}]+ [(a_{i}b_{j}-a'_{i}b'_{j})(c'_{k}d_{l}+c_{k}d'_{l})+ (a'_{i}b_{j}+a_{i}b'_{j})(c_{k}d_{l}-c'_{k}d'_{l})][(e_{m}-e'_{m})f_{n}-(e_{m}+e'_{m})f'_{n}]]Tr[\rho (\sigma_{i} \otimes \sigma_{j} \otimes \sigma_{k} \otimes \sigma_{l} \otimes \sigma_{m} \otimes \sigma_{n})].$\\
$ = \sum \limits_{i,j,k,l,m,n} [(a'_{i}b'_{j}-a_{i}b_{j})(c'_{k}d'_{l}-c_{k}d_{l})- (a'_{i}b_{j}+a_{i}b'_{j})(c_{k}d'_{l}+c'_{k}d_{l})][(e_{m}+e'_{m})f_{n}+(e_{m}-e'_{m})f'_{n}]+ [(a_{i}b_{j}-a'_{i}b'_{j})(c'_{k}d_{l}+c_{k}d'_{l})+ (a'_{i}b_{j}+a_{i}b'_{j})(c_{k}d_{l}-c'_{k}d'_{l})][(e_{m}-e'_{m})f_{n}-(e_{m}+e'_{m})f'_{n}]] t_{ijklmn}. $\\
$ = [(\overrightarrow{a'} \otimes \overrightarrow{b'}-\overrightarrow{a} \otimes \overrightarrow{b}) \otimes (\overrightarrow{c'} \otimes \overrightarrow{d'}-\overrightarrow{c} \otimes \overrightarrow{d})-(\overrightarrow{a'} \otimes \overrightarrow{b}+\overrightarrow{a} \otimes \overrightarrow{b'}) \otimes (\overrightarrow{c} \otimes \overrightarrow{d'}+\overrightarrow{c'} \otimes \overrightarrow{d})]^{T} M [(\overrightarrow{e}+\overrightarrow{e'}) \otimes \overrightarrow{f} +(\overrightarrow{e}-\overrightarrow{e'}) \otimes \overrightarrow{f'}]+ [(\overrightarrow{a} \otimes \overrightarrow{b}-\overrightarrow{a'} \otimes \overrightarrow{b'}) \otimes (\overrightarrow{c'} \otimes \overrightarrow{d}+\overrightarrow{c} \otimes \overrightarrow{d'})+(\overrightarrow{a'} \otimes \overrightarrow{b}+\overrightarrow{a} \otimes \overrightarrow{b'}) \otimes (\overrightarrow{c} \otimes \overrightarrow{d}-\overrightarrow{c'} \otimes \overrightarrow{d'})]^{T} M [(\overrightarrow{e}-\overrightarrow{e'}) \otimes \overrightarrow{f} -(\overrightarrow{e}+\overrightarrow{e'}) \otimes \overrightarrow{f'}]. $\\
Therefore from the above lemma \eqref{b3} we have,\\
$ \vert \langle S_{6} \rangle \vert \leq \lambda_{ \max } [ \vert (\overrightarrow{a'} \otimes \overrightarrow{b'}-\overrightarrow{a} \otimes \overrightarrow{b}) \otimes (\overrightarrow{c'} \otimes \overrightarrow{d'}-\overrightarrow{c} \otimes \overrightarrow{d})-(\overrightarrow{a'} \otimes \overrightarrow{b}+\overrightarrow{a} \otimes \overrightarrow{b'}) \otimes (\overrightarrow{c} \otimes \overrightarrow{d'}+\overrightarrow{c'} \otimes \overrightarrow{d}) \vert \; \vert (\overrightarrow{e}+\overrightarrow{e'}) \otimes \overrightarrow{f} +(\overrightarrow{e}-\overrightarrow{e'}) \otimes \overrightarrow{f'} \vert + \vert (\overrightarrow{a} \otimes \overrightarrow{b}-\overrightarrow{a'} \otimes \overrightarrow{b'}) \otimes (\overrightarrow{c'} \otimes \overrightarrow{d}+\overrightarrow{c} \otimes \overrightarrow{d'})+(\overrightarrow{a'} \otimes \overrightarrow{b}+\overrightarrow{a} \otimes \overrightarrow{b'}) \otimes (\overrightarrow{c} \otimes \overrightarrow{d}-\overrightarrow{c'} \otimes \overrightarrow{d'}) \vert \; \vert (\overrightarrow{e}-\overrightarrow{e'}) \otimes \overrightarrow{f} -(\overrightarrow{e}+\overrightarrow{e'}) \otimes \overrightarrow{f'} \vert].  $\\
Let us consider, $ \theta_{a} $, be the angle between $ \overrightarrow{a} $ and $ \overrightarrow{a'} $, similarly we define $ \theta_{b}\, $, $ \theta_{c}\, $  $ \theta_{d}\, $ $ \theta_{e}\, $ and $ \theta_{f} $ among the measurement directions of the parties B, C, D, E and F.\\
Using the techniques used in above Appendix \eqref{a2} we have,\\
$ \vert (\overrightarrow{e}+\overrightarrow{e'}) \otimes \overrightarrow{f} +(\overrightarrow{e}-\overrightarrow{e'}) \otimes \overrightarrow{f'} \vert^{2} = 4, $\\
$ \vert (\overrightarrow{e}-\overrightarrow{e'}) \otimes \overrightarrow{f} -(\overrightarrow{e}+\overrightarrow{e'}) \otimes \overrightarrow{f'} \vert^{2} = 4, $\\
$  \vert (\overrightarrow{a'} \otimes \overrightarrow{b'}-\overrightarrow{a} \otimes \overrightarrow{b}) \otimes (\overrightarrow{c'} \otimes \overrightarrow{d'}-\overrightarrow{c} \otimes \overrightarrow{d})-(\overrightarrow{a'} \otimes \overrightarrow{b}+\overrightarrow{a} \otimes \overrightarrow{b'}) \otimes (\overrightarrow{c} \otimes \overrightarrow{d'}+\overrightarrow{c'} \otimes \overrightarrow{d}) \vert^{2} $\\
$ = 4 (1- \cos \theta_{a} \cos \theta_{b})(1- \cos \theta_{c} \cos \theta_{d})+ 4 (1+ \cos \theta_{a} \cos \theta_{b})(1+ \cos \theta_{c} \cos \theta_{d})$.\\
$ = 8(1+ \cos \theta_{a} \cos \theta_{b} \cos \theta_{c} \cos \theta_{d} ). $\\
 Let us consider the principal angle $ \theta_{abcd}  $ such that, $  \cos \theta_{a} \cos \theta_{b} \cos \theta_{c} \cos \theta_{d} = \cos \theta_{abcd} $.\\
Hence, $  \vert (\overrightarrow{a'} \otimes \overrightarrow{b'}-\overrightarrow{a} \otimes \overrightarrow{b}) \otimes (\overrightarrow{c'} \otimes \overrightarrow{d'}-\overrightarrow{c} \otimes \overrightarrow{d})-(\overrightarrow{a'} \otimes \overrightarrow{b}+\overrightarrow{a} \otimes \overrightarrow{b'}) \otimes (\overrightarrow{c} \otimes \overrightarrow{d'}+\overrightarrow{c'} \otimes \overrightarrow{d}) \vert^{2} $\\
$ = 16 \cos^{2} \frac{\theta_{abcd}}{2} .$\\
Similarly,  $  \vert (\overrightarrow{a} \otimes \overrightarrow{b}-\overrightarrow{a'} \otimes \overrightarrow{b'}) \otimes (\overrightarrow{c'} \otimes \overrightarrow{d}+\overrightarrow{c} \otimes \overrightarrow{d'})+(\overrightarrow{a'} \otimes \overrightarrow{b}+\overrightarrow{a} \otimes \overrightarrow{b'}) \otimes (\overrightarrow{c} \otimes \overrightarrow{d}-\overrightarrow{c'} \otimes \overrightarrow{d'}) \vert^{2} $\\
$ = 16 \sin^{2} \frac{\theta_{abcd}}{2} $\\
Therefore,
   \begin{align*}
   \vert \langle S_{6}\rangle \vert  & \leq   8 \lambda_{\max} (\vert \cos \frac{\theta_{abcd}}{2}\vert + \vert \sin\frac{\theta_{abcd}}{2} \vert) \\
  & \leq 8 \sqrt{2} \lambda_{\max}. 
\end{align*}
\end{proof} 
Similar explanation also follows here as explained in Appendix \eqref{a3}.
We have consider a class of six qubit GHZ state $(\cos \phi \vert 000000 \rangle +\sin \phi \vert 111111\rangle ) $. Its optimal quantum violation is given given by
\begin{equation}
\langle S_{6} \rangle  =  8\sqrt{2} \max \lbrace 4 \sin 2\phi, 4 \sin 2\phi, 1 \rbrace. 
\end{equation}
  The optimal bound for GHZ state $ ( \phi = \frac{\pi}{4}) $ comes out to be $ 32\sqrt{2} $, maximal for the six qubit SS operator \citep{19}.
\section{Proof of Theorem. 2}\label{aa1}  
\textbf{Theorem 2.} For any four qubit local filtered state $ \rho' = \frac{1}{N'} (F_{A} \otimes F_{B} \otimes F_{C} \otimes F_{D}) \rho (F_{A} \otimes F_{B} \otimes F_{C} \otimes F_{D})^{\dagger} $ of $ \rho $ the optimal quantum bound of SS operator in Eq. \eqref{b1} is given by,
\begin{equation}\label{ll4}
V(S_{4})' = \max \vert \langle S_{4}\rangle_{\rho'} \vert\leq 4 \sqrt{2} \lambda_{\max}^{'},
\end{equation}
where $\langle S_{4}\rangle_{\rho'} = Tr(S_{4} \rho')$, and $ \lambda'_{\max} $ is the maximal singular value of the matrix $ \frac{\Lambda}{N'} $, where $ \Lambda $ is defined in Eq. \eqref{lr3} taking over all quantum states, that are locally unitary equivalent to $ \rho $. Consequently, $ \lambda'_{\max} $  is also the maximal singular value of the matrix $ M' = [t'_{ijkl}] $, where $ t'_{ijkl} = Tr [ \rho' (\sigma_{i} \otimes \sigma_{j} \otimes \sigma_{k} \otimes \sigma_{l})],\; i,j,k,l = 1,2,3$.
\begin{proof}
The normalization factor $ N' $ is given by,\\
\begin{align*}
 N' & = Tr [ ( K \Sigma^{2}_{A} K^{\dagger} \otimes  L \Sigma^{2}_{B} L^{\dagger} \otimes  M \Sigma^{2}_{C} M^{\dagger} \otimes  N \Sigma^{2}_{D} N^{\dagger}) \rho] \\
 & = Tr[ (\Sigma^{2}_{A} \otimes \Sigma^{2}_{B} \otimes \Sigma^{2}_{C} \otimes \Sigma^{2}_{D})\\
 & ( K^{\dagger} \otimes L^{\dagger} \otimes M^{\dagger} \otimes N^{\dagger})\rho (K \otimes L \otimes M \otimes N)]\\
 & = Tr[ (\Sigma^{2}_{A} \otimes \Sigma^{2}_{B} \otimes \Sigma^{2}_{C} \otimes \Sigma^{2}_{D}) \varrho],  
\end{align*}
where we have considered $ \varrho = ( K^{\dagger} \otimes L^{\dagger} \otimes M^{\dagger} \otimes N^{\dagger})\rho (K \otimes L \otimes M \otimes N) $. Since the two states $ \rho $ and $ \varrho $ are locally unitary equivalence, hence they have the same value with respect to the maximum violation of SS inequality.\\
From the double cover relationship \cite{54,55} between the special unitary group $ SU(2) $ and the special orthogonal group $ SO(3) $, $ K \sigma_{i} K^{\dagger} = \sum_{j=1}^{3} O_{ij} \sigma_{j} $, where $ K $ is any given unitary operator and the matrix $ O $ with entries $ O_{ij} $ belongs to $ SO(3) $, we have \\
\begin{equation}\label{l6}
\begin{split}
 & t'_{ijkl} \\
& = Tr[\rho' (\sigma_{i} \otimes \sigma_{j} \otimes \sigma_{k} \otimes \sigma_{l})]\\
&= \frac{1}{N'} Tr[(F_{A} \otimes F_{B} \otimes F_{C} \otimes F_{D}) \rho (F_{A} \otimes F_{B} \otimes F_{C} \otimes F_{D})^{\dagger}\\
& (\sigma_{i} \otimes \sigma_{j} \otimes \sigma_{k} \otimes \sigma_{l})]\\
& = \frac{1}{N'} Tr[\rho(K \Sigma_{A}K^{\dagger}\sigma_{i} K \Sigma_{A}K^{\dagger} \otimes L \Sigma_{B}L^{\dagger}\sigma_{j} L \Sigma_{B}L^{\dagger}\\
&  \otimes M \Sigma_{C} M^{\dagger}\sigma_{k} M \Sigma_{C}M^{\dagger} \otimes N \Sigma_{D}N^{\dagger}\sigma_{l} N \Sigma_{D}N^{\dagger})]\\
& = \frac{1}{N'} \sum \limits_{p,q,r,s} Tr[(K^{\dagger}\otimes L^{\dagger}\otimes M^{\dagger}\otimes N^{\dagger}) \rho (K \otimes L\otimes M \otimes N)\\
&(\Sigma_{A} O^{A}_{ip} \sigma_{p}\Sigma_{A} \otimes \Sigma_{B} O^{B}_{jq} \sigma_{q}\Sigma_{B}\otimes \Sigma_{C} O^{C}_{kr} \sigma_{r}\Sigma_{C} \otimes \Sigma_{D} O^{D}_{ls} \sigma_{s}\Sigma_{D})]\\
&=\frac{1}{N'} \sum \limits_{p,q,r,s} O^{A}_{ip}O^{B}_{jq}O^{C}_{kr}O^{D}_{ls}*\\
& Tr[\varrho(\Sigma_{A} \sigma_{p}\Sigma_{A} \otimes \Sigma_{B} \sigma_{q}\Sigma_{B}\otimes \Sigma_{C} \sigma_{r}\Sigma_{C} \otimes \Sigma_{D} \sigma_{s}\Sigma_{D})] \\
&= \frac{1}{N'} \sum \limits_{p,q,r,s} O^{A}_{ip}O^{B}_{jq}O^{C}_{kr}O^{D}_{ls} Tr[\varrho (\alpha_{p} \otimes \beta_{q}\otimes \gamma_{r} \otimes \delta_{s})]\\
& = \frac{1}{N'} \sum \limits_{p,q,r,s} O^{A}_{ip}O^{B}_{jq}O^{C}_{kr}O^{D}_{ls} \Lambda_{pqrs}\\
& =  \frac{1}{N'} [(O_{A} \otimes O_{B})\Lambda (O^{T}_{C} \otimes O^{T}_{D})]_{ijkl}
\end{split}
\end{equation}
Hence, we have $ M' = \frac{1}{N'} [(O_{A} \otimes O_{B})\Lambda (O^{T}_{C} \otimes O^{T}_{D})] $, where
\begin{equation}\label{l7}
\begin{split}
 M'^{\dagger}M' & = \frac{1}{N'^{2}} [ (O_{C}\otimes O_{D}) \Lambda^{\dagger} (O^{T}_{B} \otimes O^{T}_{A})*\\
 & (O_{A} \otimes O_{B}) \Lambda (O^{T}_{C} \otimes O^{T}_{D})]\\
  & = \frac{1}{N'^{2}}[ (O_{C}\otimes O_{D}) \Lambda^{\dagger} \Lambda (O^{T}_{C} \otimes O^{T}_{D})]
  \end{split}
 \end{equation} 
 Since the operators $ (O_{C}\otimes O_{D}) $ are orthogonal, consequently $  M'^{\dagger}M' $ has the same eigenvalues of $ \dfrac{\Lambda^{\dagger}\Lambda}{N'^{2}} $. Hence $ M' $ has the same singular values of $ \dfrac{\Lambda}{N'} $. If $ \overrightarrow{v} $ is a nine dimensional singular vector of $ \dfrac{\Lambda}{N'} $ then $ (O_{C}\otimes O_{D})\overrightarrow{v} $ is that of $ M' $.
\end{proof}

\end{document}